\documentclass[aps,prb,reprint,showkeys,showpacs]{revtex4-2}
\pdfoutput=1
\usepackage{amsmath}
\usepackage{amssymb}
\usepackage{amsfonts}
\usepackage{graphicx}
\usepackage{bm}
\usepackage{siunitx}
\usepackage{color}
\usepackage{hyperref}
\hypersetup{%
    pdfborder = {0 0 0}
}

\allowdisplaybreaks
\usepackage{cleveref}

\graphicspath{{./figures/}}

\renewcommand\vec{\bm} 
\newcommand{\uGrad}{\vec{\nabla}}

\newcommand{\ud}{\,{\mathrm{d}}}

\newcommand{\uiiint}{\int\!\!\!\int\!\!\!\int}

\newcommand{\uC}{C} 
\newcommand{\uD}{D} 
\newcommand{\ue}{e} 
\newcommand{\uE}{E} 
\newcommand{\uF}{F} 

\newcommand{\uHopf}{{\cal{H}}} 
\newcommand{\uh}{h} 
\newcommand{\uH}{H} 
\newcommand{\uvH}{\vec{H}}     
\newcommand{\uK}{K}         

\newcommand{\uvM}{\vec{M}}
\newcommand{\uvm}{\vec{m}}
\newcommand{\umi}{m_i}      
\newcommand{\umx}{m_{\mathrm{X}}}
\newcommand{\umy}{m_{\mathrm{Y}}}
\newcommand{\umz}{m_{\mathrm{Z}}}

\newcommand{\uO}{O}         
\newcommand{\uP}{P_i}
\newcommand{\uPex}{P_\mathrm{EX}}

\newcommand{\uPdm}{P_\mathrm{DM}}

\newcommand{\up}{p_i}
\newcommand{\upex}{p_\mathrm{EX}}
\newcommand{\upa}{p_\mathrm{A}}
\newcommand{\upz}{p_\mathrm{Z}}
\newcommand{\updm}{p_\mathrm{DM}}
\newcommand{\upIdm}{p^\mathrm{I}_\mathrm{DM}}
\newcommand{\upIIdm}{p^\mathrm{II}_\mathrm{DM}}

\newcommand{\uMs}{M_{\mathrm{S}}}

\newcommand{\umuZ}{\mu_0}
\newcommand{\uq}{q} 
\newcommand{\uvs}{\vec{s}}  
\newcommand{\uX}{X}         
\newcommand{\uY}{Y}         
\newcommand{\uZ}{Z}         
\newcommand{\uXp}{\widetilde{X}} 
\newcommand{\uYp}{\widetilde{Y}} 
\newcommand{\uZp}{\widetilde{Z}} 
\newcommand{\uR}{R}    
\newcommand{\ur}{r}
\newcommand{\urp}{\widetilde{r}}
\newcommand{\uvr}{\vec{r}}          
\newcommand{\uvrp}{\widetilde{\vec{r}}}  

\begin{document}
\title{Elliptical stability of hopfions in bulk helimagnets}

\author{Konstantin L. Metlov}\email{metlov@donfti.ru}
\affiliation{Galkin Donetsk Institute for Physics and Engineering, R.~Luxembourg str.~72, Donetsk 283048, Russian Federation}

\date{\today}
\begin{abstract}
Magnetic hopfions are three-dimensional topological solitons with non-zero Hopf index ${\cal H}$ in the vector field of material's local magnetization. Here elliptical stability of hopfions with ${\cal H}=1$ in a classical helimagnet is studied on the basis of a variational model. It is shown that, depending on their internal structure (vortex and antivortex tubes ordering), the hopfions can either be stable in a bulk magnet or unstable with respect to elongation along their central axis. It is found that the energy of stable hopfions is always below the energy of the $2\pi$-skyrmion lattice in the same material, suggesting the possibility to use $2\pi$-skyrmions as a precursor for hopfion nucleation. Stability diagram for hopfions on the magnetic anisotropy-field phase diagram is computed numerically. Explicit analytical expressions for some of its critical lines are derived.
\end{abstract}


\maketitle

\section{Introduction}
Topological objects are ubiquitous in magnetism. These are one-dimensional (1D) domain walls~\cite{Hubert_Shafer}, which may also acquire a complex two-dimensional (2D) structure~\cite{HSG1958,M01_CT};  2D skyrmions in bulk helimagnets~\cite{BY1989,*BH1994} and thin films~\cite{bobeck1975bubbles}, magnetic vortices~\cite{UP93} in planar nanostructures~\cite{M10}. In fact, setting the coherent rotation of the magnetization aside -- motion, pinning and resonant dynamics of the topological objects to large extent define the static (hysteretic) and dynamic properties of magnetic media. The common ground between all of these topological objects is that, assuming periodic boundary conditions in space, all of them correspond to mappings of a sphere to a sphere. The target sphere is always $S^2$ --- the set of endpoints of the 3D magnetization vector $\uvM$ of the fixed length $|\uvM(\uvr)|=\uMs$, while the source sphere describes the topology of space: in the 1D case it is a circle $S^1$, in the 2D case~\cite{BP75} it is a Riemann sphere $S^2$. Such sphere to sphere mappings ($S^1\rightarrow S^2$ and $S^2\rightarrow S^2$) split into integer-numbered homotopy classes. Any magnetization distribution of the relevant dimensionality can be classified by computing the corresponding integer --- the topological index (topological charge).

In three dimensions, topological solitons correspond to maps $S^3\rightarrow S^2$ with $S^3$ being a sphere with 3D surface, embedded in four-dimensional space. Originally in mathematics it was accepted that all such mappings are homotopically equivalent to each other, until Heinz Hopf in 1931 provided a counter-example~\cite{Hopf1931}. It was then generalized by Whitehead~\cite{whitehead1947}, who classified all $S^3\rightarrow S^2$  mappings by an integer topological index (Hopf invariant). Magnetic hopfions, embedded in the magnetization vector field of a bulk magnet, were hypothesized by Dzyaloshinskii and Ivanov~\cite{DI79} long ago. But only recent progress in bulk 3D nanoscale imaging techniques~\cite{CMSGHBRHCG2020} opens a way for experimental observation of magnetic hopfions, which are regarded as a key ingredient of the emerging 3D nano-magnetism~\cite{Gubbiotti2024roadmap}. However, important questions remain: Which materials can support hopfions and at what conditions ? How to create them ?

Topological aspects of hopfions are firmly established since the work of Whitehead~\cite{whitehead1947}: if a magnetization distribution is spatially localized and has the Hopf index $\uHopf>0$ --- it is a hopfion. But their energetic aspects, required to answer the above fundamental questions, are still poorly understood. Stable hopfions and hopfion-like states were obtained numerically, constrained by the sample geometry~\cite{TS2018,sutcliffe2018,lake2018,WQB2019} or other magnetic patterns in the sample~\cite{Voinescu2020,Kuchkin2023}, and experimentally~\cite{kent2021,Yu2023,zheng2023}. Yet, these are not free standing bulk hopfions and their existence depends on the external stabilization and confinement. Usually the latter is done by sandwiching the hopfion-containing film between two magnetic layers with strong perpendicular magnetic anisotropy~\cite{TS2018}. However, the common experience from simulations is that the hopfion size increases with distance between the confining layers, implying that the hopfions are laterally stable (in the film plane), but elliptically unstable.

\section{Model}
The starting point is the micromagnetic energy of a classical helimagnet per unit volume $\uE=\lim_{V\rightarrow\infty} (1/V) \uiiint_V F \ud^3\uvr$ with the density
\begin{align}
        \uF = &\frac{\uC}{2}\sum_{i=\uX,\uY,\uZ} \left|\uGrad\umi\right|^2 +\uD\, \uvm\cdot[\uGrad\times\uvm] - \nonumber \\ 
        &\umuZ \uMs \left(\uvm \cdot \uvH\right) - \uK\left(\uvm \cdot \uvs\right)^2, 
        \label{eq:energy} 
\end{align}
where $\uvm(\uvr)=\uvM(\uvr)/\uMs$, $|\uvm|=1$, $\uC=2A$ is the exchange stiffness in \si{\joule\per\meter}, $\uD$ is the Dzyaloshinskii-Moriya (DM) interaction strength~\cite{BS1970,BJ1980} in \si{\joule\per\square\meter}, $\uK$ is the uniaxial anisotropy constant in \si{\joule\per\cubic\meter} and $\uvs$ is its director, $\uvH$ is the external magnetic field strength in \si{\ampere\per\meter} and $\umuZ$ is the permeability of vacuum. For simplicity, let's assume that the field and the anisotropy axis are parallel and choose Cartesian coordinate system in such a way that both are directed along the $\uO\uZ$ axis: $\uvH=\{0,0,H\}$, $\uvs=\{0,0,1\}$. Our task now is to find $\uvm(\uvr)$, which minimizes $\uE$ and has $\uHopf>0$.

The exact solution to this problem is unknown. But, as it is usual in micromagnetics, an approximate solution can be found assuming a parametrized magnetization distribution and minimizing its energy with respect to those parameters. Provided the energy is computed exactly (as in the present work), such a Ritz-type~\cite{Ritz09} approach yields an upper bound for the (unknown) exact solution energy of the considered variational problem. The lower this upper bound estimate is for a particular Ritz-type model, the closer it is to the exact solution energy. In this work we start with:
\begin{subequations}
\label{eq:model}
\begin{align}
\label{eq:stereogr}
&\{\umx+\imath\umy,\umz\}=\{2w, 1 - |w|^2\}/(1+|w|^2),\\
\label{eq:whitehead}
&w=\pm\imath u/v,\\
\label{eq:E3S3}
 &u=\frac{2(\uXp+\imath \uYp)\uR}{\urp^2+\uR^2},
 \ 
 v=\frac{\uR^2-\urp^2+\imath 2 \uZp \uR}{\urp^2+\uR^2},
\\
\label{eq:physE3}
&\uvrp=\uR\,\frac{\uvr}{r}\frac{\ue(r/\uR)}{1-\ue(r/\uR)}, \  \uvr=\{\uX,\uY,\gamma\uZ\},
\end{align}
\end{subequations}
$r=|\uvr|$, $\uvrp=\{\uXp,\uYp,\uZp\}$, $\urp=|\uvrp|$. It defines $\uvm(\uvr)$, parametrized by an unknown hopfion profile function $\ue(x)$, satisfying the boundary conditions $\ue(0)=0$, $\ue(1)=1$, and two scalar parameters: the hopfion radius $\uR$ and the aspect ratio $\gamma$. The hopfion is fully contained within the ellipsoid $r<1$ with the magnetization $\uvm=\{0,0,1\}$ on the hopfion boundary ($r=1$) and outside. When $|\gamma|<1$ the hopfion is elongated along $\uO\uZ$ direction. General design of~\eqref{eq:model} is discussed in detail in~\cite{M2023_TwoTypes,M2024ms}. Briefly, it consists of the stereographic projection~\eqref{eq:stereogr} mapping a complex number $w$ (belonging to a Riemann sphere $S^2$) to the magnetization vector with guaranteed unit length, the Whitehead's~\cite{whitehead1947} ansatz~\eqref{eq:whitehead} mapping the sphere $S^3$ in double complex parametrization ($|u|^2+|w|^2=1$) to the Riemann sphere $w$ with $\uHopf=1$ and an additional helicity prefactor (one of the two equilibrium values~\cite{M2023_TwoTypes} with top and bottom signs for Type~I and Type~II hopfions respectively), map~\eqref{eq:E3S3} of the extended Euclidean space $\uvrp$ to the sphere $S^3$ and, finally, the map~\eqref{eq:physE3} of an interior of the ellipsoid $r<1$ in physical space to the extended Euclidean space (such that the boundary of the ellipsoid is mapped to the infinitely distant point).

There are two distinctions from the earlier work~\cite{M2023_TwoTypes,M2024ms}. The first is an addition of the aspect ratio parameter $\gamma$. The $\uHopf=1$ hopfion's magnetization distribution (see
\begin{figure}[tb]
\begin{center}
\includegraphics[trim={0.18cm 0.3cm 0.18cm 0.18cm},width=\columnwidth]{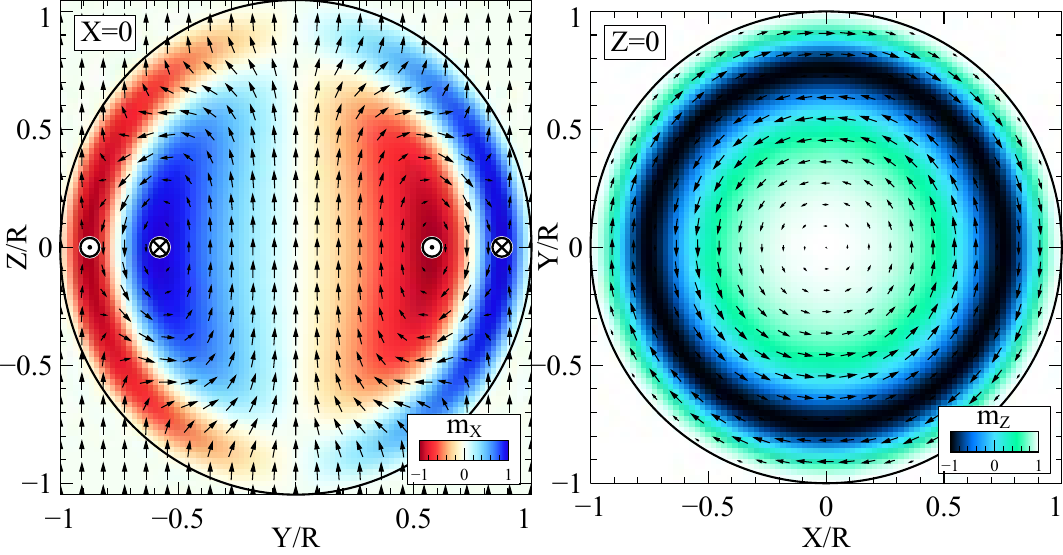}
\end{center}
\caption{\label{fig:H1}Cross sections of the equilibrium hopfion~\eqref{eq:model} at $H=0$, $K=0$ by the $X=0$ (left) and $Z=0$ (right) planes. Chirality of this trial function corresponds to the negative equilibrium values~\cite{M2023_TwoTypes} of the Type~I hopfion size parameter $\nu$ in the Eq.~\eqref{eq:energyF} or to $\uD<0$. For $D>0$ chirality of the trial function needs to be switched (e.g. by reversing the sign of $\uY$ and $\umy$).}
\end{figure}
Fig.~\ref{fig:H1}) is axially symmetric and $\gamma$ allows to scale it along the symmetry axis. A very similar hopfion structure (except for the overall compression along the $\uO\uZ$ axis due to confinement) was obtained numerically for $\uD>0$ in~\cite{sutcliffe2018}. Spherical harmonics expansion, used to evaluate the hopfion's magnetostatic energy in~\cite{M2023_TwoTypes,M2024ms}, works well to factor the hopfion's magnetostatic energy into the radial and angular parts (see the Appendix C in~\cite{M2024ms}), which allows to integrate the latter analytically and greatly simplifies numerical treatment of the considered variational problem. Unfortunately, this trick can not be directly applied to elliptically deformed hopfions, requiring a different set of Laplace operator eigenfunctions, matching the hopfion's ellipsoidal boundary. For this technical reason, the dipolar interaction is not considered here. On the other hand, it merely increases the hopfion's energy, compared to (mostly) pole-free helimagnet ground states, and destabilizes spherical hopfions~\cite{M2024ms}. Stability region of elliptical hopfions, computed in the present work, can be expected to evolve in a similar fashion with magnetostatic interaction strength.

The second difference lies in the parametrization of the physical space mapping~\eqref{eq:physE3}. Formerly~\cite{M2023_TwoTypes,M2024ms} it was specified in terms of the function $f(x)$ as $\uvrp=\uvr/(1-f(r/R))$ with the boundary conditions $f(0)=0$, $f'(0)=0$, $f(1)=1$. The difference might seem minor, but it is not! It alone is responsible for reducing the energy of the equilibrium hopfion at $H=0$, $K=0$ from $\uE\uC/\uD^2=0.56506-0.60162=-0.03656$ for the model~\cite{M2023_TwoTypes} to $0.2198-0.4397=-0.2199$ in the present model. Where the total energies are split into the exchange (positive) and Dzyaloshinskii-Moriya (negative) contributions. For comparison, the energy of the conical ground state at the same conditions is $0.5-1=-0.5$. The reason for such a dramatic energy reduction is that the function $f(x)$ was overconstrained by the unnecessary boundary condition $f(0)=0$ (the condition $f'(0)=0$ is still required to ensure continuity of the magnetization vector derivatives at $\uvr=0$). Its removal allows to reach a much deeper energy minimum. Note that this removal is {\em all that sufficient} and the minimization can still be done in terms of the function $f(x)$ with the corresponding physical space mapping. Reparametrization to $0\le e(x)\le1$ is more of a convenience, ensuring that the profile always stays bounded (unlike $f(x)$, which now becomes unbounded at $0$). The two functions are related by $f(x)=1+x-x/\ue(x)$ and the condition $f'(0)=0$ is equivalent to $\ue''(0)=-2[\ue'(0)]^2$, which is automatically satisfied if $\ue(x)$ is the solution of the Euler-Lagrange equation minimizing $\uE$.

\section{Equilibrium energy}
Substituting the trial function~\eqref{eq:model} into~\eqref{eq:energy} and assuming that hopfions form a close-packed 3D lattice (FCC or HCP) with each one occupying the volume $V=4\sqrt{2}R^3\gamma$, the total energy per unit volume can be expressed as $\uE\uC/\uD^2=\int_0^1{\cal F}\ud x$ with
\begin{equation}
 \label{eq:energyF}
 {\cal F}= \frac{1}{\kappa} \left[\nu^2\upex+\nu\,\updm+\frac{q}{2}(\upa-\kappa)+h(\upz-\kappa)\right],
\end{equation}
where $\kappa=4\sqrt{2}$, $\nu=\uC/(\uD R)$ is the dimensionless (inverse) size parameter, $\uh=\umuZ\uMs\uC\uH/\uD^2$ is the normalized external field, $\uq=2\uC\uK/\uD^2$ is normalized anisotropy quality factor and the energy function integrands $\up$ are given in the Appendix~\ref{a:pi}. The energy $\uE$ is a functional of the hopfion profile function $e(x)$ with two additional scalar parameters --- $\nu$ and $\gamma$.

Let us briefly discuss the dependence on $\gamma$. There are two hopfion types~\cite{M2023_TwoTypes}, corresponding to different signs in~\eqref{eq:whitehead}. The Type~I hopfion (shown in Fig.~\ref{fig:H1}) consists of an outer anti-vortex tube wound on top of the inner vortex tube. In Type~II hopfions (with ``minus'' sign in~\eqref{eq:whitehead}) the tube order is reversed.  In this work, the change of the hopfion type can also be achieved by changing the sign of $\gamma$. It mirrors the hopfion with respect to the $\uZ=0$ plane, which turns the vortices into antivortices and vice-versa. It can be directly shown that $\upex(\gamma)=\upex(-\gamma)$ is the same for hopfions of either type~\cite{M2024ms}, while the DM energy of the Type~I hopfions $\upIdm=\updm(\gamma)$ turns into the negative DM energy of the Type~II hopfions upon the $\gamma$ sign reversal $\upIIdm=-\updm(-\gamma)$. This $\updm$ sign change compensates the change of the $\nu$ sign  between the hopfions of different types~\cite{M2023_TwoTypes}. The other energies --- $\upa$ and $\upz$ are independent on $\gamma$ and on the hopfion type. Thus, by extending the range of $\gamma$, hopfions of both types can be considered within the same framework. With top sign in~\eqref{eq:whitehead} the positive $\gamma$ corresponds to the ellipsoidal Type~I hopfions and the negative to the Type~II hopfions with the aspect ratio $|\gamma|$.

Computation of the equilibrium hopfion profile is convenient to express as a boundary value problem for a system of ordinary differential equations (ODEs). To this end we first convert the scalar parameters into functions, by assuming $\nu\equiv\nu(x)$ and $\gamma\equiv\gamma(x)$ and introduce two new unknown functions $\uE_\nu(x)=\int_0^x\partial{\cal F}/\partial\nu \ud x$ and $\uE_\gamma(x)=\int_0^x\partial{\cal F}/\partial\gamma \ud x$, such that $\uE_\nu(1)=\partial\uE/\partial\nu$ and $\uE_\gamma(1)=\partial\uE/\partial\gamma$. Then, finding the equilibrium hopfion at given $q$ and $h$ (the only two remaining external parameters) reduces to solving the following 6-th order system of ODEs and boundary conditions:
\begin{subequations}
\label{eq:minimize}
\begin{align}
 &\frac{\partial {\cal F}}{\partial e} - \frac{\partial}{\partial\ur}\frac{\partial {\cal F}}{\partial e'} = 0,\  e(0)=0,\  e(1)=1 \\
 & \uE_\nu'(x)=\frac{\partial {\cal F}}{\partial \nu},\  \nu'(x)=0,\  \uE_\nu(0)=0,\  \uE_\nu(1)=0\\ 
 & \label{eq:Egamma}\uE_\gamma'(x)=\frac{\partial {\cal F}}{\partial \gamma},\  \gamma'(x)=0,\  \uE_\gamma(0)=0,\  \uE_\gamma(1)=0,
\end{align}
\end{subequations}
which can be conveniently done with the shooting method. For numerical results here the {\tt NDSolve} function of Wolfram Mathematica\texttrademark{} was used.

\section{Elliptical stability of hopfions}
To study the elliptical stability of hopfions, let us temporarily forget about the equations~\eqref{eq:Egamma} and solve the remaining 4-th order ODEs considering $\gamma$ an external parameter. The resulting hopfions have equilibrium profiles $e(x)$ and an equilibrium size $\nu$ for a particular hopfion's aspect ratio. The energy of such hopfions, as function of $\gamma$ for several different values of $q$ and $h$ is plotted in 
\begin{figure}[tb]
\begin{center}
\includegraphics[trim={0.18cm 0.3cm 0.18cm 0.18cm},width=\columnwidth]{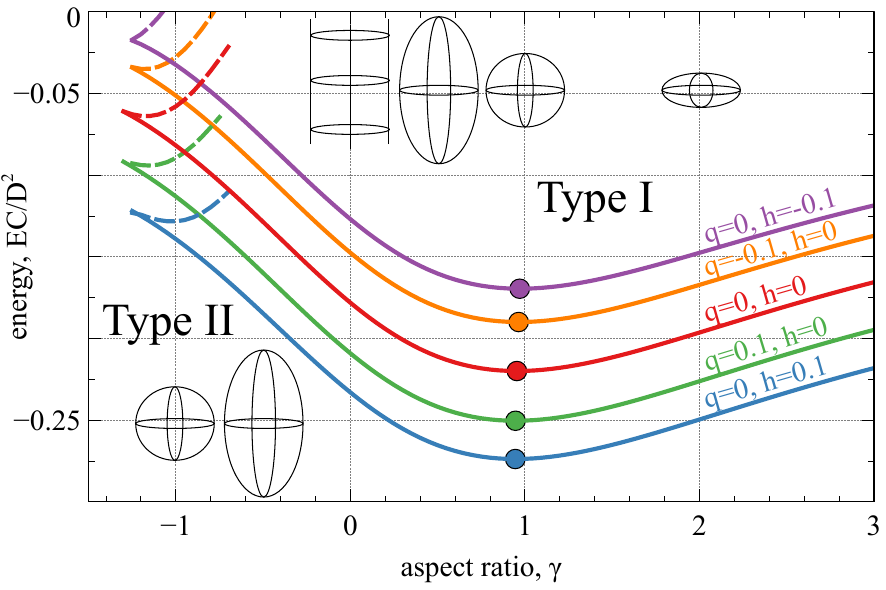}
\end{center}
\caption{\label{fig:eofgamma}The equilibrium hopfion energy as function of the aspect ratio parameter $\gamma$ for several combinations of the anisotropy $q$ and the magnetic field $h$ around $q=h=0$. Dashed lines on the left, show the energy maxima in $\nu$ parameter (for each combination of $q$ and $h$), which neighbor the energy minima, shown by the solid lines. Sketches visualize the shape of hopfions for different $\gamma$. Circles mark positions of the global energy minima. At $q=h=0$ and $\gamma=\pm1$ the energy $\uE\uC/\uD^2$ is $0.2198-0.4396=-0.2198$ and $0.082-0.164=-0.082$ for spherical Type~I and Type~II hopfions respectively.}
\end{figure}
Fig.~\ref{fig:eofgamma}. It shows clearly that the Type~I hopfions are elliptically stable. They are mostly spherical ($\gamma\approx1$), but do almost always have some degree of elliptical deformation. The Type~II hopfions are elliptically unstable. If created, they expand in the $\uO\uZ$ direction, become columnar at $\gamma=0$  and then convert to the Type~I hopfions. However, this expansion can be stopped artificially (e.g. by sandwiching the hopfion-containing film between two pinning layers with higher uniaxial anisotropy).

But what about the smaller values of $\gamma<-1$, may be there is an energy minimum for the Type~II hopfions there? The answer is no. To prove it, in addition to the energy at the minimum over $\nu$, the energy at the neighboring maximum in $\nu$ was computed and plotted in Fig.~\ref{fig:eofgamma} by dashed lines. At the leftmost end of the shown curves, the maxima and minima merge, implying that the energy minimum in $\nu$ turns into an inflection point. Thus, further squeezing (increase of $|\gamma|$) destroys hopfion's lateral stability. One can also see that the Type~I hopfions are much more stable with respect to squeezing. Interestingly, a similar generalization by adding the aspect ratio $\gamma$ to the older hopfion model~\cite{M2023_TwoTypes,M2024ms} results in two energy minima: one at positive and another at negative gamma, implying the elliptical stability of Type~II hopfions as well. This artifact is removed in the present model due to a more thorough energy minimization.

One may note that the energy goes smoothly through $\gamma=0$, corresponding to the infinite cylindrical hopfions. Such hopfions consist of repetition of the magnetization distribution in the hopfion's $Z=0$ plane (right side of the Fig.~\ref{fig:H1}) for every $Z$. This suggests that such a two-dimensional ring domain structure (also called  $2\pi$-skyrmion lattice) can be used as a precursor for nucleating hopfions. Once a ring domain state is created, it could (if not pinned by the interfaces or defects) spontaneously relax to the lower energy Type~I hopfion.

Hopfion stability can be studied numerically, by finding points in the $q$--$h$ parameter space, where the solution to~\eqref{eq:minimize} ceases to exist. This was done by starting with a stable hopfion solution at $h=q=0$ and tracing along the rays in all possible directions with successively decreasing steps. As soon as {\tt NDSolve} ceases to find a solution along such a ray despite the step (distance to the previous stable point, used to provide the initial solution estimates) decreased below a certain very small threshold, the last stable point is marked. These points are shown in
\begin{figure}[tb]
\begin{center}
\includegraphics[trim={0.2cm 0.45cm 0.25cm 0.25cm},width=\columnwidth]{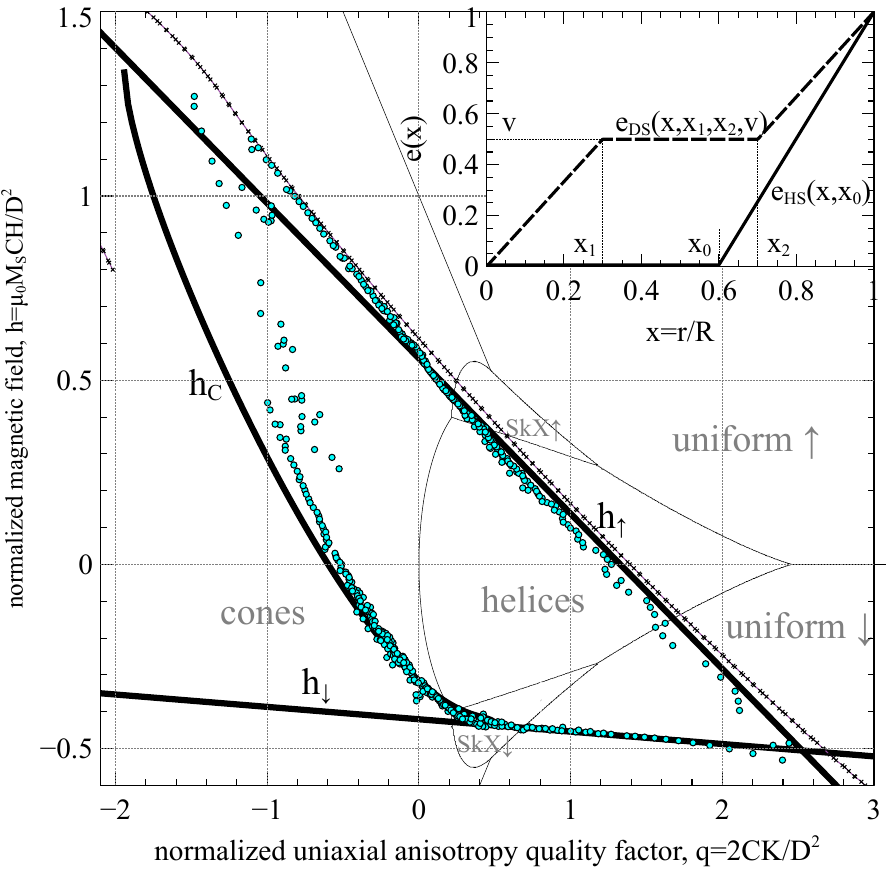}
\end{center}
\caption{\label{fig:stabilityDiagram}Stability region of the hopfion state, superimposed over the classical ground state diagram of a helimagnet. Circles show the numerical results~\cite{M2025figshare} from the model~\eqref{eq:model}, crosses show an approximation to the upper stability line using the spherical hopfion model~\cite{M2023_TwoTypes}, the lines $h_\uparrow$, $h_\downarrow$ and $h_\mathrm{C}$ are computed analytically in the text. Inset shows trial functions for the hopfion profile, used in the analytical calculation.}
\end{figure}
Fig.~\ref{fig:stabilityDiagram}, superimposed over the classical helimagnet ground state diagram (consisting of the uniform, conical, helical and skyrmion phases). The stability region of the present hopfion model with elliptical deformation is smaller, compared to the earlier spherical model~\cite{M2023_TwoTypes}. This happens everywhere, except the high-field stability line, where the numerical results from the spherical (shown by crosses in Fig.~\ref{fig:stabilityDiagram}) and the elliptical hopfion models nearly coincide. 

The top line marks the second order transition to the uniform phase, where the hopfion energy becomes equal to the energy of the uniform magnetization (below this line the hopfion energy is lower). The transition happens via the expansion of the hopfion core (nearly uniform region around $X=Y=0$) until the rest of the hopfion structure (vortex and antivortex tubes) disappears. While the hopfion always has some degree of elliptical deformation inside its stability region, on approach to the critical lines the equilibrium aspect ratio $\gamma$ tends to unity. At the critical lines the hopfion size $\uR\propto1/\nu$ either expands to infinity or collapses to zero and in both cases the spherical hopfion shape is optimal. It also happens that at the top stability line the computed equilibrium $f(0)$ becomes almost equal to zero. That's why the spherical hopfion model~\cite{M2023_TwoTypes} reproduces it rather well.

To obtain analytical approximation to this line, consider a simple trial function in the form of the hockey stick (see the inset in Fig.~\ref{fig:stabilityDiagram})
\begin{equation}
 \label{eq:eHS}
 e_\mathrm{HS}(x, x_0) = \frac{x-x_0}{1-x_0}\theta(x-x_0),
\end{equation}
where $x_0$ is the hopfion core size (in units of $R$) and $\theta(x)$ is the Heaviside theta function [$\forall x >0: \theta(x)=1, \theta(-x)=0$]. Its energy $\uE_\mathrm{HS}(\nu, x_0)$ can be computed by substituting~\eqref{eq:eHS} into~\eqref{eq:energyF} and integrating. It is quadratic in $\nu$ and can be minimized analytically by setting $\nu=\nu_\mathrm{eq}=-\uPdm/(2\uPex)$, where $\uP=\int_0^1\up\ud x$. The stability line results from the condition $\ud\uE_\mathrm{HS}(\nu_\mathrm{eq}, x_0)/\ud x_0 = 0$ at $x_0=1$ that the vortex core completely engulfs the whole hopfion and $x_0=1$ becomes stable. With $\gamma=1$ it is a straight line, shown in Fig.~\ref{fig:stabilityDiagram}:
\begin{equation}
\label{eq:hup}
 h_{\uparrow}=\frac{4 (2+\pi ) (3 \pi -4) q-15 \pi ^2}{60 (\pi -4) (2+\pi )}.
\end{equation}

The transition to the uniform state, magnetized oppositely to the hopfion core, can be considered using the trial function with two linear slopes and the horizontal line in the middle (also shown in the inset on Fig.~\ref{fig:stabilityDiagram})
\begin{align}
 \label{eq:eDS}
 e_\mathrm{DS}=&\frac{v x\!+\!v (x_1\!-\!1) e_{\mathrm{HS}}(x,x_1)\!-\!(v\!-\!1) x_1 e_{\mathrm{HS}}(x,x_2)}{x_1}
\end{align}
where $v$ is the value of $e_\mathrm{DS}$ in the horizontal part $x_1\le x\le x_2$. It describes instability around a particular point of the $e(x)$ map, whose neighborhood becomes so energetically favorable that it starts to expand, ultimately collapsing the rest of the hopfion. In the $Z=0$ plane the point with $\umz=-1$ corresponds to  $e(x)=1/2$. Thus, to consider the (start of the) conversion to $\umz=-1$ state, we compute $\uE$ with $e=e_\mathrm{DS}$,$v=1/2$, $\nu=\nu_\mathrm{eq}$, $\gamma=1$ and check (see the Appendix~\ref{a:eDS} for details) the stability with respect to enlargement of the uniform region (when $x_1$ starts to be different from $x_2$). This yields:
\begin{equation}
 \label{eq:hdown}
 h_{\downarrow}=\frac{4 (2+\pi ) (3 \pi -10) q-15 \pi ^2}{60 \left(\pi ^2-4\right)},
\end{equation}
plotted in Fig.~\ref{fig:stabilityDiagram}. It compares very well to the numerical results almost everywhere in the $q>0$ region, except the region of $q\lesssim0.6$. This is because at these $q$ the ground state becomes conical, instead of uniform.

To study the transition into conical state the same $e_\mathrm{DS}$ model can be used with $v$ chosen in such a way that the angle between the magnetization and the $\uO\uZ$ axis in the $Z=0$ plane is the same as the conical angle (in the conical state). This calculation (described in the Appendix~\ref{a:eDS}) gives the $h_\mathrm{C}$ stability line, shown in Fig.~\ref{fig:stabilityDiagram}. It has  an excellent agreement with the numerically computed points at $q>0$ with worse agreement at negative $q$.

Of course, with $e_\mathrm{DS}$ a particular value of $v$ only guarantees a particular $\uvm$ direction in the $Z=0$ plane of the hopfion. Above and below the plane the direction at the same value of $e$ will be different. Yet, finding the weakest point is sufficient for the stability analysis. The functions $e_\mathrm{HS}$ and $e_\mathrm{DS}$ lead to very precise $h_\downarrow$ and $h_\mathrm{C}$ lines at $q>0$, which should suffice for practical estimates of hopfion stability in real materials.

In the case of the easy-plane anisotropy both the numerically computed stability lines (which are visibly jagged) and the analytical ones are in much worse agreement. On the other hand, the model~\eqref{eq:model} with the uniform background, is bound to significantly overestimate the hopfion energy in the easy-plane magnet. While a better estimates for the stability lines of the present model for $q<0$ can be sought, it is, probably, a more pressing concern to improve the model~\eqref{eq:model} itself in this region.

One should also have in mind that by its nature any Ritz-type model reduces the number
of degrees of freedom of the considered variational problem to make it tractable. Such models are well suited to study particular instability modes (such as elliptical instability, considered here), but can not provide a proof of the absolute stability of a particular magnetization configuration. Yet, usually there are not many instability modes and a Ritz type model can evolve to cover them all, as it happened in many well known micromagnetic problems.

The lowest energy hopfions are located around the $h_\uparrow$ line in the first ($h>0$, $q>0$) quadrant of the phase diagram. That's where the present analytical estimates are very precise. They are also good  at $h<0$, $q>0$ for hopfions, created at $h>0$ with subsequent field reversal.

Let as also note that the above piecewise linear function models for $e(x)$ are only good at the fringes of the hopfion's stability region. Inside, a much better analytical models for $e(x)$ can be built, which will be a subject of forthcoming publication.

\section{Conclusions}
Elliptical stability of magnetic hopfions in a classical helimagnet is studied using a variational model. Depending on the hopfion type (defined by its internal structure), the hopfions can either be elliptically stable (when their antivortex filament is wound atop of the vortex filament) or elliptically unstable (when the order of filaments is reversed) with the tendency to expand indefinitely in the direction of the hopfion's central axis. In the latter case the perpendicular magnetic anisotropy modulation in layers, orthogonal to the hopfion axis, can stop the expansion and stabilize the hopfions. It is shown that equilibrium hopfions always have a lower energy than the $2\pi$-skyrmion (concentric ring domain) lattice in the same material, which suggests that $2\pi$-skyrmions can be used as a precursor for controlled hopfion creation. The variational model for the hopfion profile is improved and the corresponding phase diagram, taking into account possible elliptical deformation, is computed numerically. For two of the stability lines in this phase diagram the explicit analytical expressions are obtained and the approximation to the remaining line is computed implicitly.

This work was supported by the Russian Science Foundation under the project RSF 25-22-00076.

\appendix
\section{Hopfion energy function integrands}
\label{a:pi}
For brevity let's omit the arguments of the profile function:  $\ue=\ue(x)$, $\ue^\prime=\ue^\prime(x)$, so that the brackets in the expressions below are only used for grouping. Then the exchange energy function integrand is
\begin{align}
\label{eq:pex}
 \upex = & 
 \frac{64 \pi  \left(10 p^2+x^2 (5   + 3 \cos \alpha)\right)(e^\prime)^2}{15 (1-2p)^2 (\cos\alpha+1)} +
 \nonumber\\
 & \frac{256 \pi  p (1-2 e) (2 p (1-2 e- x e^{\prime})+ x e^{\prime}) \cos\alpha}{15 (1-2p)^4 (\cos\alpha+1)},
\end{align}
where $\gamma=\tan(\alpha/2)$ and $p=e(1-e)$. For spherical hopfions $\gamma=\pm1$ ($\alpha=\pm\pi/2$) this expression coincides with $\upex$ in~\cite{M2024ms}, converted to the $e(x)$ profile function. It is also evident that the exchange energy is an even function of $\gamma$.

For the Dzyaloshinskii-Moriya integrand we have
\begin{align}
\label{eq:pdm}
 \updm = & \frac{32 \pi x^2 e^\prime (5 - 4 p (5-p(2 \gamma +3)))}{15 (1-2p)^3} + \nonumber\\
 & \frac{64 \pi x p (1-2 e) (5-4 p (5-p(4 \gamma +1)))}{15 (1-2p)^4},
\end{align}
for $\gamma=1$ it coincides with $\upIdm$ and for $\gamma=-1$ with $-\upIIdm$ from~\cite{M2024ms}, converted to the $e(x)$ profile function.

The anisotropy and Zeeman integrands are independent of $\gamma$ and are identical to those of the spherical hopfions, sans the conversion to the $e(x)$ profile function:
\begin{align}
\label{eq:pa}
\upa = &\frac{128 \pi x^2 p^2 (5-4p (5-p))}{15 (1-2p)^4}, & \\
\label{eq:pz}
\upz = &\frac{64 \pi x^2 p^2}{3 (1-2p)^2}.
\end{align}

\section{Computing stability lines using the double-slope trial function}
\label{a:eDS}
The energy per unit volume $\uE$ of the hopfions with the profile $e_{DS}$ is straightforward to compute analytically by substituting~\eqref{eq:eDS} into~\cref{eq:pex,eq:pdm,eq:pa,eq:pz} as $e$, then to~\eqref{eq:energy} and integrating over $x$. Also assuming (as confirmed by the numerical calculation) that the hopfions near the critical state are nearly spherical we can set $\gamma=1$. The hopfion energy is then a function of five dimensionless parameters: $\uE(h,q,v,\overline{x},\Delta x)$, where we have defined $\overline{x}=(x_1+x_2)/2 \in [0,1]$ and $\Delta x=(x_2-x_1)/2\in [0,1/2]$. While the expression of this function is too voluminous to include here, it is still very easily manageable using a computer algebra system.

To study stability towards the conversion into the uniform state, opposite to the hopfion core, we further let $v=1/2$ and expand the normalized energy density by assuming $\overline{x}=1/2+\alpha\delta$ and $\Delta x=1/2-\delta$ for $0<\delta\ll1$:
\begin{align}
 \uE\uC/\uD^2 =&\frac{2}{45} \sqrt{2} \pi  (5
   h+q)-h-\frac{q}{2} + \nonumber \\
   &\frac{\pi F(h,q)(\alpha -1)\delta}{90 \sqrt{2} (2+\pi )}+O(\delta^2)
\end{align}
with $F=60 (\pi ^2-4) h+3 \pi ^2 (5-4 q)+16 \pi  q+80 q$. The condition $F(h_\downarrow,q)=0$ then results in~\eqref{eq:hdown}. It means that the state with the uniform profile $e(x)\equiv1/2$ had just became stable and the hopfion had vanished. Note that the condition~\eqref{eq:hdown} essentially means that the vortex and antivortex cores of the hopfion got erased from its $\uZ=0$ plane. Conversion of the rest of the hopfion to the uniform state is sure to follow, but its detailed consideration is beyond the present variational model.

To consider transition into a conical state, we use the same expression for $\uE$ as above, but select $v$ in such a way that it corresponds to a equilibrium projection of the magnetization onto the field in a conical state, that is $m_z=h/(1-q)$. There are two solutions to the resulting quadratic equation
\begin{equation}
 v=\frac{1}{2} \left[1\mp\sqrt{\frac{2 \sqrt{2} \sqrt{(q\!-\!1) (h\!+\!q-\!1)}\!+\!h\!+\!3 q\!-\!3}{-\!h\!+\!q\!-\!1}}\right]
\end{equation}
of which the one with the $-$ sign is chosen, because it leads to the least stable hopfion. Unfortunately, this time it is impossible to find the explicit solution for the stability line, which is defined by the transcendental system of equations
\begin{equation}
 \left.\frac{\partial\uE}{\partial\overline{x}}\right|_{\Delta x=0}=0,\ 
 \left.\frac{\partial\uE}{\partial\Delta x}\right|_{\Delta x=0}=0,
\end{equation}
defining a line $h=h_\mathrm{C}(q)$, corresponding to the start of disappearance of the energy minimum at $\Delta x=0$ or, in other words, to the start of the expansion of the region of $e(x)=v$ from a point at $x=\overline{x}$ to a finite range of $x$. The line $h=h_\mathrm{C}(q)$ is also plotted in Fig.~\ref{fig:stabilityDiagram}.

%
\end{document}